\documentclass[aip,reprint]{revtex4-1}
\usepackage{graphicx}
\usepackage{listings}
\usepackage{picins}
\usepackage{latexsym}
\usepackage{amsmath}
\usepackage{booktabs}
\usepackage{threeparttable}

\draft 

\begin{document}


\title{Double lattice potential for molecular dynamics simulation of silicon with demonstrated validity} 



\author{Hui Zhang}
\email[Electronic mail:]{zhope@scut.edu.cn}
\noaffiliation
\author{Chongyang Wei}
\noaffiliation
\author{Zhongwu Liu}
\noaffiliation
\author{Xichun Zhong}
\noaffiliation
\author{Dongling Jiao}
\noaffiliation
\author{Wanqi Qiu}
\noaffiliation
\author{Hongya Yu}
\noaffiliation
\affiliation{School of Materials Science and Engineering, South China University of Technology, Guangzhou 510640, People's  Republic of China}


\date{\today}
\begin{abstract}
To reproduce the diamond structure of silicon, double lattice (DL) potential constructed from two interatomic potentials for face centered cubic (fcc) lattice, is proposed for molecular dynamics (MD) simulations. For the validity test of MD simulation, the Tersoff potential, the Stillinger and Weber (SW) potential, the environment-dependent interatomic (EDI) potential, the charge optimized many-body (COMB) potential, and the modified embedded-atom (MEAM) potential have been also employed for comparison. The crystal lattice of simulated silicon system is identified by calculating the distribution functions of the distances between the atoms and the angles between the lines linking an atom with its nearest neighbors. The results are also compared with the perfect silicon crystal. The crystal lattice, the crystallization temperature, and elastic constants have been calculated from MD simulations using above potentials. The results show that the systems with modified Tersoff, SW, EDI, COMB, and MEAM potentials could not exhibit the diamond structure and only the DL potential gives diamond lattice. The ground state for DL potential is the wurtzite structure, and the metastable state formed during rapid cooling is the cubic diamond structure. The physical parameters obtained from the simulation with DL potential are in agreement with the experiment results. This work indicated that only DL potential is valid for MD simulation of silicon crystal among above various potentials.
\end{abstract}


\maketitle 
\section{Introduction}
Silicon is a technologically important semiconductor and one of the most heavily studied materials \cite{Schulz-1,Dmitrienko-1,Remsing-1,Darkins-1,Dorner-1,Eliassen-1,Yokoi-1,Bartok-1}. During past decades, atomistic computation methods have been employed to study the structures of Si and its compounds along with the experimental methods \cite{Martin-1}. With the decrease in silicon chip size, the experimental investigations on both physical and chemical behaviors of the nano silicon clusters become more and more difficult and the atomistic computation methods become especially important. However, silicon is a covalent crystal and the description for the covalent bond in atomistic computation methods is a hard issue. Theoretically, the density functional theory (DFT) can give the most accurate result, but the time and the computation power required for the system consisting of a few hundred of atoms are huge. A feasible alternative approach is to construct an empirical potential, and then to calculate both the physical and chemical parameters of the system by molecular dynamics (MD) simulations. 

So far, the interatomic potentials for silicon mainly include Tersoff potential \cite{Tersoff-1,Tersoff-2} and its modified versions \cite{Kumagai-1,Pun-1}, Stillinger and Weber (SW) potential \cite{Stillinger-1}, environment-dependent interatomic (EDI) potential \cite{Bazant-1}, charge optimized many-body (COMB) potential \cite{Yu-1}, and modified embedded-atom (MEAM) potential \cite{Baskes-1} and its modified version \cite{Lenosky-1}. Among these, Tersoff potential, SW potential, EDI potential, and COMB potential are many-body potentials. In these potentials, in addition to the pair term for the interatomic coupling between two atoms, there is a three-body term that can describe the interatomic coupling among three atoms or define the so-called bond angles. Furthermore, COMB potential also takes into account the effects of charge. It is expected that these many-body potentials could present an appropriate description for silicon. To test their validity, the researchers \cite{Tersoff-1,Tersoff-2,Kumagai-1,Pun-1,Stillinger-1,Bazant-1,Yu-1,Baskes-1,Lenosky-1} have calculated the relevant parameters based on above different potentials, which includes the elastic constants, binding energy, radial distribution function, and so on. The test of physical parameters is important, but what is more important is whether the diamond structure could be reproduced by the simulation using these interatomic potentials. 

In fact, it was found that the diamond structure could not be formed in MD simulations even if the physical parameters given with one interatomic potential are in good agreement with the experimental data. For instance, the embedded-atom (EAM) potential is successfully employed for the metallic elements and alloys \cite{Daw-1}, and the experiment results of crystal lattices could be reproduced by MD simulations \cite{Zhou-1,Mendelev-1}. However, for the silicon showing diamond structure, with the above-mentioned interatomic potentials, no simulated systems can form the diamond structure in MD simulations. Thus, these potentials could not pass the validity test of crystal structure. Therefore, the objective of this work is to construct a new interatomic potential for silicon, with which the diamond structure could be reproduced by MD simulation and the physical parameters can be obtained.

The diamond structure could be regarded as the penetration of two face centered cubic (fcc) lattices with eight atoms in its crystal cell. It is thus reasonable to reproduce the diamond structure from two fcc interatomic potentials and one interatomic potential between two lattices. So far there mainly existing potential for fcc lattice is EAM potentials \cite{Daw-1,Zhou-1,Mendelev-1}, but too many parameters introduced in EAM potentials make it difficult to define the interatomic coupling between two fcc lattices. Here, we only take into account the simple Lenard-Jones (LJ) potential. Our recent investigation has shown that the ground state of LJ potential corresponds to the hexagonal close packed (hcp) lattice ($c/a\approx1.633$, $a$ and $c$ being lattice constants of hcp lattice) by MD simulation, and the systems could also show the meta-stable fcc lattice under some simulation circumstances \cite{Zhang-1}. The Bravais lattices have been identified by comparing the simulation results with those of perfect hcp and fcc lattices. Further investigations have indicated that with the combination of several LJ potentials and without setting any initial Bravais lattices, the diamond and graphite structures \cite{Zhang-2}, and more complex perovskite ABO$_{3}$ structure \cite{Zhang-3} can be reproduced in MD simulations. These results are helpful for understanding and rethinking the significance of LJ potential and the formation of crystal structure.

In this work, double lattice (DL) potential has been constructed to reproduce the diamond structure of silicon. DL potential and the above-mentioned other interatomic potentials have been employed for MD simulation. The crystal structure, the crystallization temperature, lattice constants, and elastic constants have been obtained and compared with the experiment results.

\section{Modeling and Simulations}
\subsection{The modeling}
As mentioned above, a double lattice potential, named DL potential, is constructed in this work, and the basic idea is to describe the diamond structure by two fcc interatomic potentials and one interatomic potential between two lattices. The simple LJ potential used for describing fcc lattice can be expressed as:
\begin{eqnarray}
U(r)=4\epsilon \left(\left(\frac{\sigma}{r}\right)^{12}-\left(\frac{\sigma}{r}\right)^{6}\right)
\end{eqnarray}
where $\epsilon$ is the depth of the potential well, $\sigma$ is the distance at which the potential is zero, and $r$ is the distance between the atoms.

Two types of atoms denoted as S$_{1}$ and S$_{2}$ are introduced in the model. They are physically identical to each other, and may correspond to two spin configurations. For an example, there are four spins up for S$_{1}$ atom and four spins down for S$_{2}$ atom. S$_{1}$ and S$_{2}$ atoms should show their own lattices, but their lattices and lattice constants are the same. Here, unlike other models in which all Si atoms are the same, S$_{1}$ and S$_{2}$ atoms in our model are treated as two distinguishable Si atoms. There are interactions between S$_{1}$ and S$_{2}$ lattices. Here, LJ potential is used for S$_{1}$ lattice, S$_{2}$ lattice, and the interaction between S$_{1}$ and S$_{2}$ lattices. The equilibrium distance between the atoms and the crystallization temperature are determined by LJ potential.
\begin{table}[h]
\centering
\caption{The parameters for LJ potentials in MD simulations }
\begin{tabular}{p{12mm}p{12mm}p{12mm}p{12mm}}
\hline
\toprule
Atom & $\epsilon$(eV) & $\sigma$(\AA)	 & $r_{c}$(\AA)\\
\hline
\midrule
S$_{1}$   &   0.22   &   3.58   &   3$\sigma$\\
S$_{2}$ & 0.22 & 3.58 &  3$\sigma$\\
S$_{1}$S$_{2}$ & 0.22 & 1.94 &  3$\sigma$\\
\bottomrule
\hline
\end{tabular}
\end{table}
\subsection{Simulation detail}
In the simulation, without setting any initial Bravais lattices, S$_{1}$ and S$_{2}$ atoms are equally and randomly created in the simulation box with the boundary conditions applied. For LJ potential, the cutoff is denoted as $r_{c}$. The maximum temperature for the system is not higher than the temperature $T_{0}$ (4000 K). The NPT ensemble is annealed at $T_{0}$ for 1000 picosecond with a timestep of $10^{-3}$ picosecond. The temperature of the system $T$ decreases from $T_{0}$ with a temperature step $n$, and then the NPT ensemble is annealed for 100-1000 picosecond at each temperature $T$. The pressure is zero in all simulations. The parameters for LJ potentials in MD simulation are listed in Table I. 

To obtain the stress versus strain relations for Si crystal from MD simulations, a perfect Si crystal is created. The initial fcc lattices for S$_{1}$ and S$_{2}$ atoms have the same lattice constants of 5.341 \AA  \, and the relative displacements of S$_{1}$ lattice with respect to S$_{2}$ lattice are 0.25, 0.25, and 0.25 in the  $x$, $y$, and $z$ directions, respectively. The NPT ensemble is first annealed for 1000 picosecond at 300 K, and then small deformations of the system follow. The values for the stress caused by the deformations are calculated and the stress versus strain relation is obtained. Consequently, the elastic constants $c_{11}$, $c_{12}$, and $c_{44}$ are calculated and the in-script can be referred to Ref. [25]. 

In this work, different potentials mentioned earlier such as Tersoff, SW, EDI, COMB, and MEAM potentials are also employed for MD simulation comparison. The physical parameters from DL potential are compared with those from other potentials. The numbers of atoms for the simulations are 512, 1000, and 8000. MD simulations are carried out with the aid of lammps software \cite{Plimpton-1} , and the visualization is done with VESTA software \cite{Momma-1}.

\subsection{Identification of the crystal structure of the system}
To test the validity of DL potential, the most important thing to do is to check whether the system exhibits the diamond structure. Here the distribution functions of distances between atoms and the angles between the lines linking an atom with its nearest neighbors, i.e. $\rho(d)$ and $\rho(\theta)$, respectively, are calculated for the simulated system. A reference system of perfect silicon crystal is obtained from lammps software. The difference between simulated and the reference systems is checked for the lattice identification. For perfect Si crystal, the distribution functions consist of a small number of discrete values. If the distribution functions from the simulations show non-zero values at these corresponding positions, then the system can be considered to have the same crystal structure as the perfect Si crystal. In the simulation, if the coordinate of any atom at any temperature at any time ($x$, $y$, $z$) is known so that both the distances between atoms and the angles between the lines linking an atom with its nearest neighbors can be calculated. The distribution functions  $\rho(d)$(or $\rho(\theta)$) mean the count values or intensity of the distances $d$ (or the angles $\theta$) in the range of $d$-$d$+d$d$ (or $\theta$-$\theta$+d$\theta$). In the calculation, the distances between one atom $R$ and its nearest neighbors are defined as $d_{R}$. Theoretically, the values for $d_{R}$ in the system are the same but they show some difference in MD simulation. Here, we denote $d_{m}$ as the minimum distance of these distances $d_{R}$. As a result, the atom is regarded as one of the nearest neighbors of atom $R$ if $d_{R}/d_{m} < 1.1$.

In addition, the distribution function $\rho(d)$ is similar to the conventional radial distribution function $g(r)$, but their algorithms are different. The angle $\theta$ is the bond angle for some crystal structures like the diamond structure. In the case of silicon, for the calculation of distribution function, S$_{1}$ and S$_{2}$ atoms can be treated as one type of atoms for the whole system, and also can be treated separately for S$_{1}$ (S$_{2}$) sub-system.
 \begin{figure}[h t b p]
\centering
 \includegraphics[width=80mm]{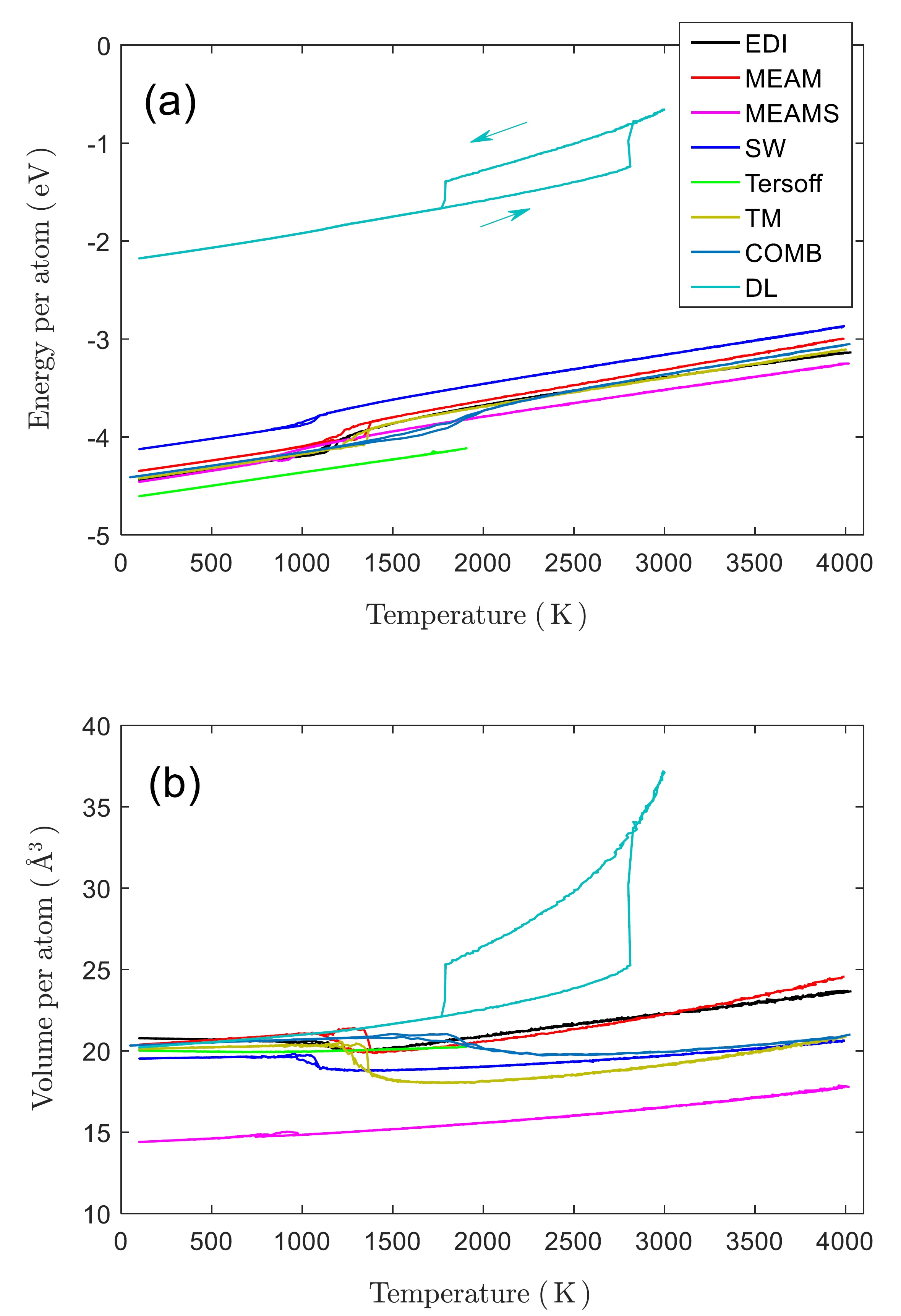}%
 \caption{\label{FIG.1.}Dependences of energy per atom (a) and volume per atom (b) on the temperature with different interatomic potentials applied in MD simulations. Here, MEAMS stands for MEAM-Spline potential, and TM for Tersoff-Mod potential. In (a), the arrows show the directions for the decrease and increase in temperature, respectively.}
 \end{figure}
 \begin{figure}[h t b p]
\centering
 \includegraphics[width=70mm]{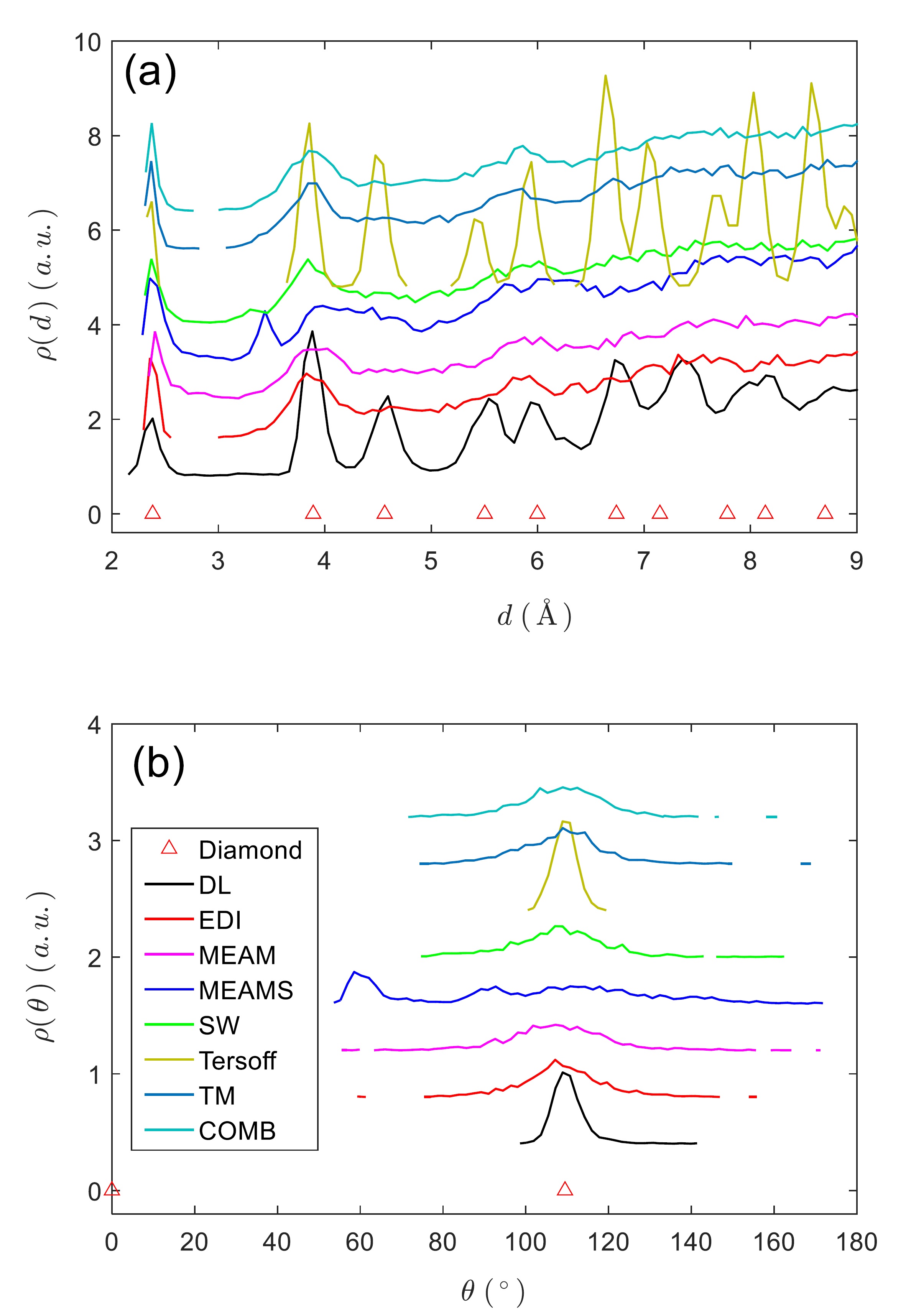}%
 \caption{\label{FIG.2.}The distribution functions of the distances between atoms and the angles between the lines linking one atom with its nearest neighbors $\rho(d)$ (a) and $\rho(\theta)$ (b) with different interatomic potentials applied in MD simulations at  $T$=100 K. For COMB potential $T$=50 K. Note that the wurtzite structure has the same $\rho(d)$ and $\rho(\theta)$ as the diamond structure.}
 \end{figure}
 \begin{figure}[h t b p]
\centering
 \includegraphics[width=70mm]{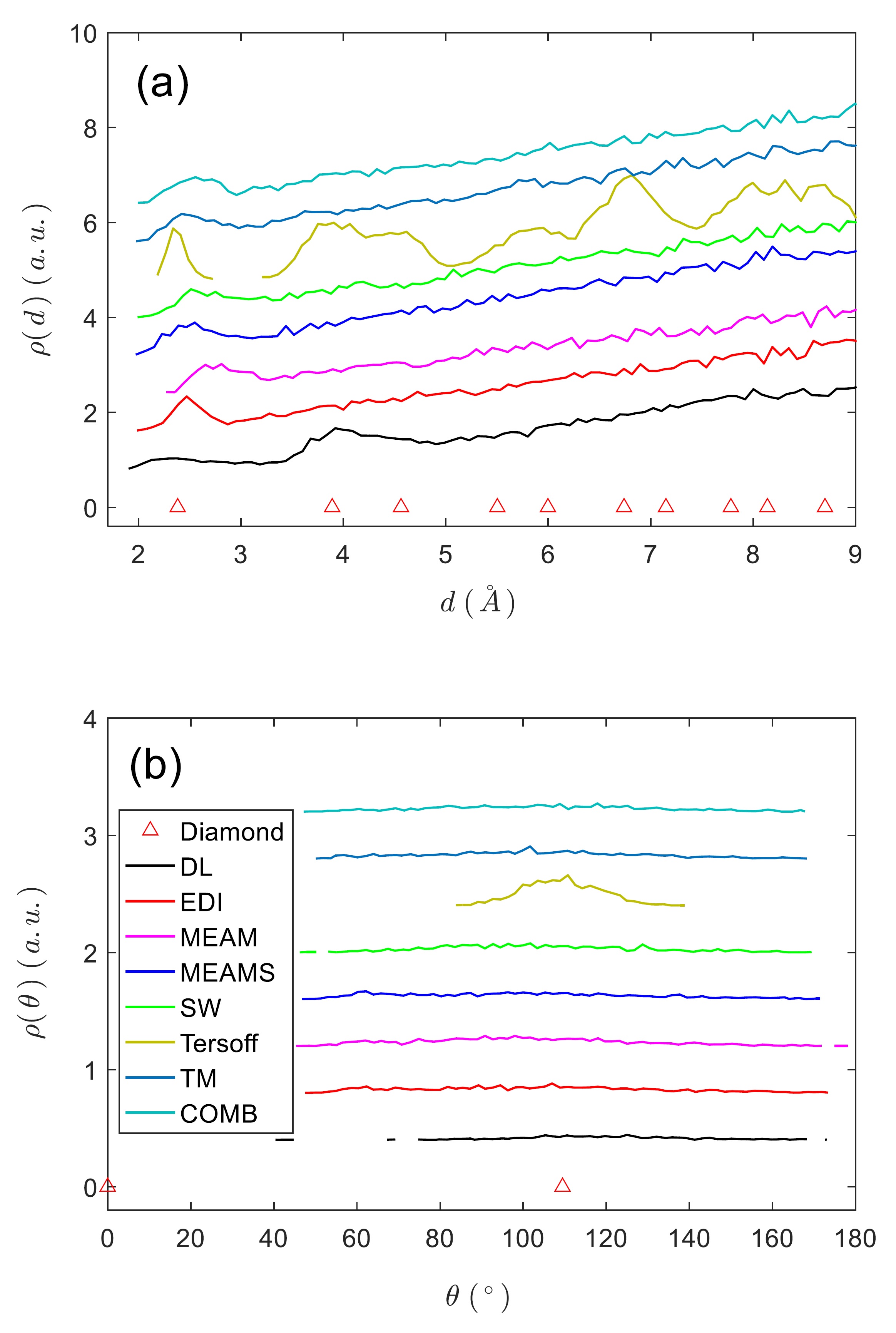}%
 \caption{\label{FIG.3.}The distribution functions of the distances between atoms and the angles between the lines linking one atom with its nearest neighbors $\rho(d)$ (a) and $\rho(\theta)$ (b) with different interatomic potentials applied in MD simulations at $T$=4000 K. For Tersoff potential $T$=1800 K  and for DL potential $T$=3000 K.}
 \end{figure}
\begin{figure*}
\centering
 \includegraphics[width=140mm]{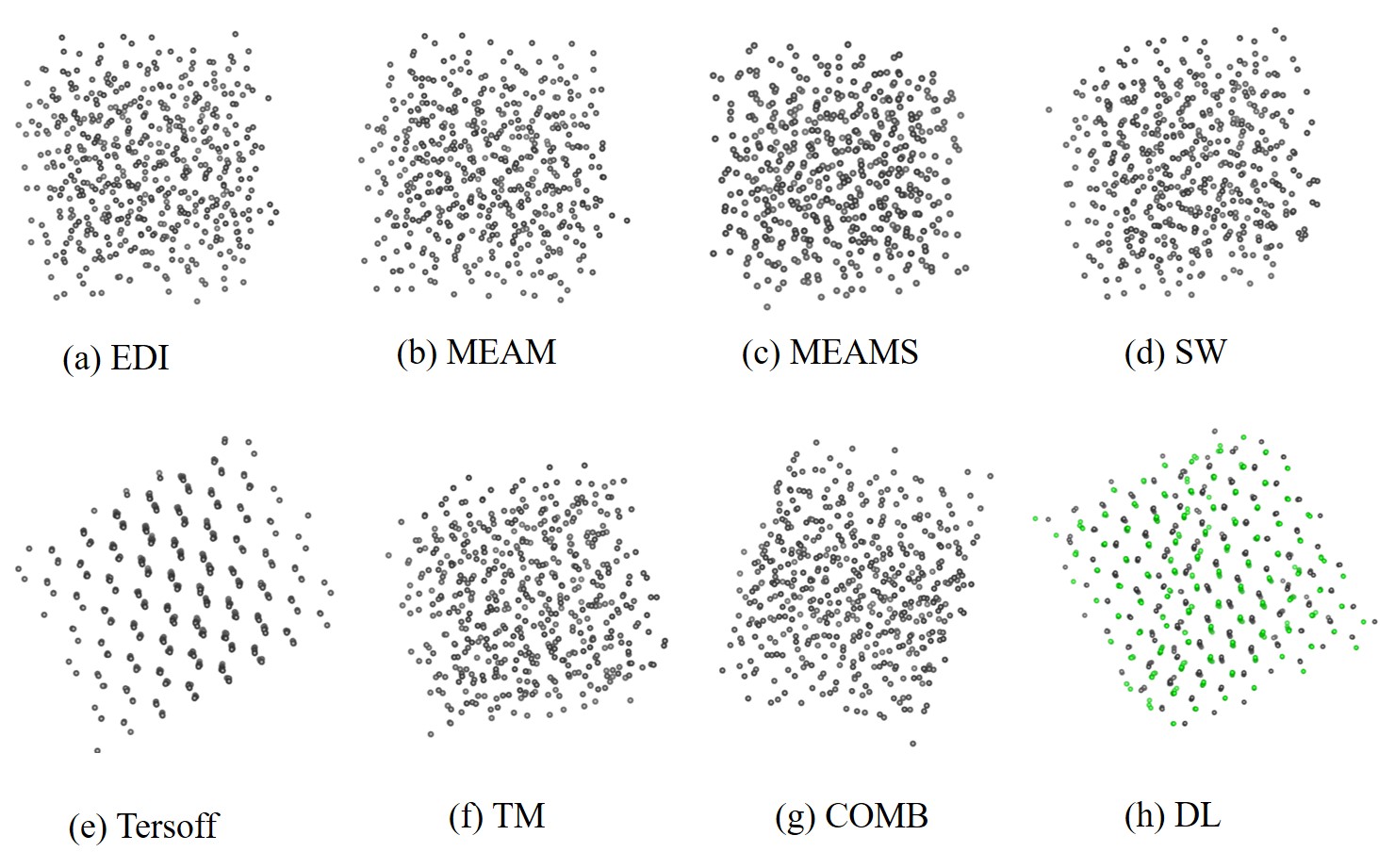}%
 \caption{\label{FIG.4.} The atomic configurations with different interatomic potentials applied in MD simulations at $T$=100 K. For COMB potential $T$=50 K. In (h), S$_{1}$ and S$_{2}$ Si atoms are indicated by black and green dots, respectively.}
 \end{figure*}
 \begin{figure}[h t b p]
\centering
 \includegraphics[width=70mm]{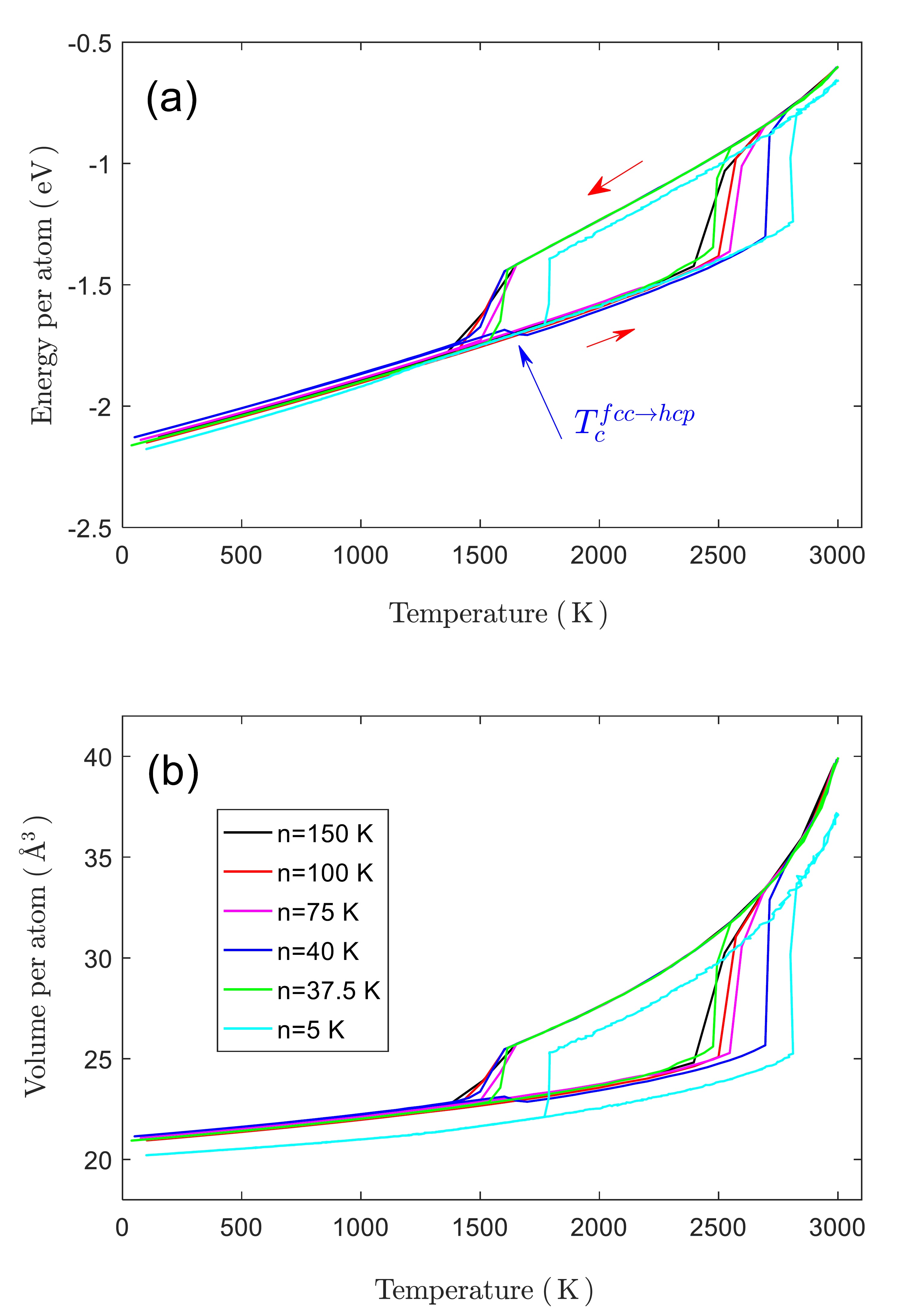}%
 \caption{\label{FIG.5.}Dependence of energy per atom (a) and volume per atom (b) on the temperature with different temperature steps $n$ in MD simulations for DL potential.}
 \end{figure}
    \begin{figure*}
 \centering
 \includegraphics[width=140mm]{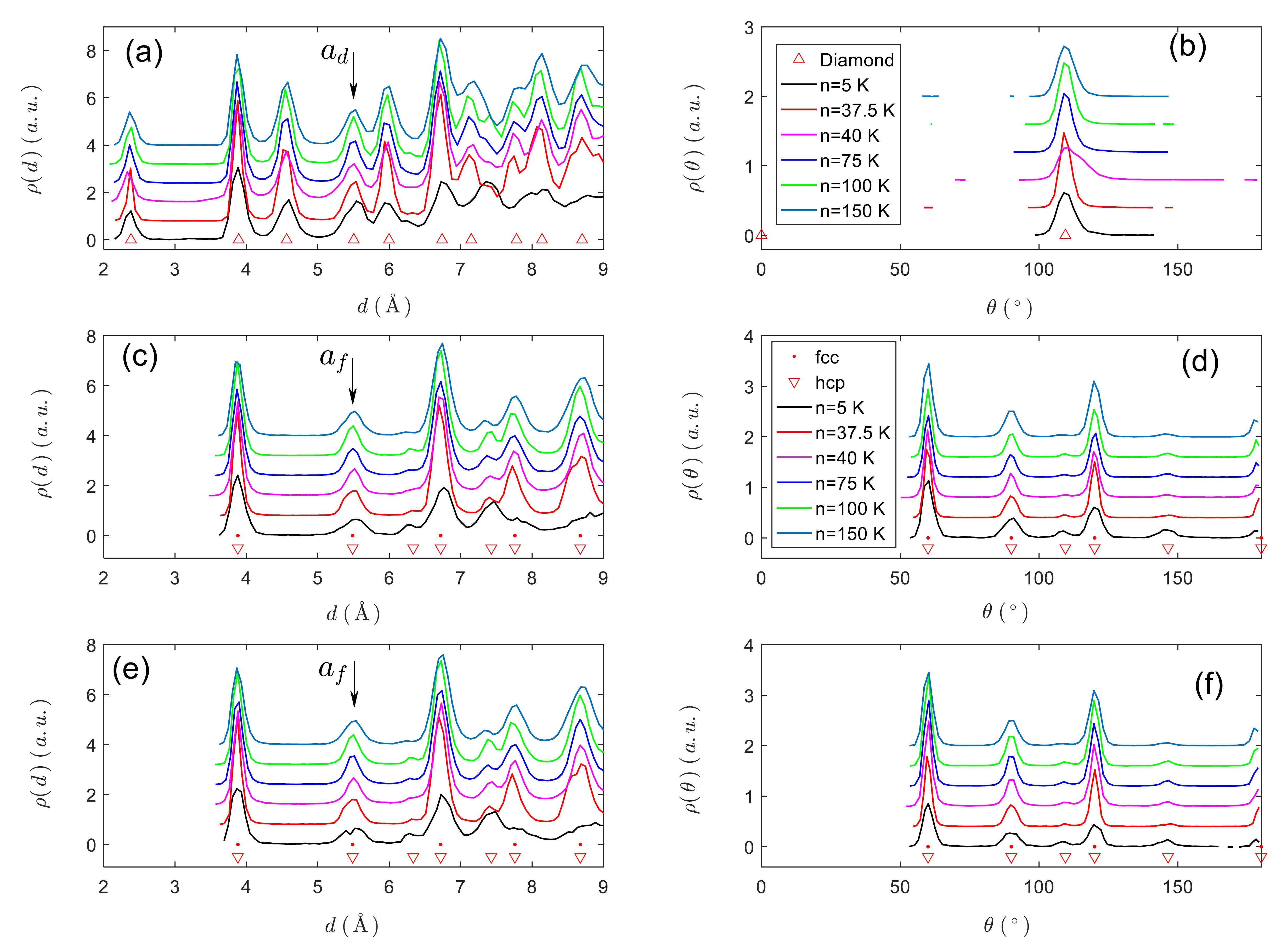}%
 \caption{\label{FIG.6.}The distribution functions of the distances between atoms and the angles between the lines linking one atom with its nearest neighbors $\rho(d)$ and $\rho(\theta)$ with different temperature steps $n$ applied in MD simulations at low temperatures for DL potential. (a)-(b) are for the whole S$_{1}$+S$_{2}$ system, and (c)-(d) and (e)-(f) for S$_{1}$ and S$_{2}$ subsystem, respectively. In (a), (c), and (e), $a_{d}$ and $a_{f}$ stand for the lattice constants for the diamond structure and fcc lattice, respectively, as indicated by the arrows. Note that the wurtzite structure has the same $\rho(d)$ and $\rho(\theta)$ as the diamond structure.}
 \end{figure*}
 \begin{figure}[h t b p]
 \centering
 \includegraphics[width=70mm]{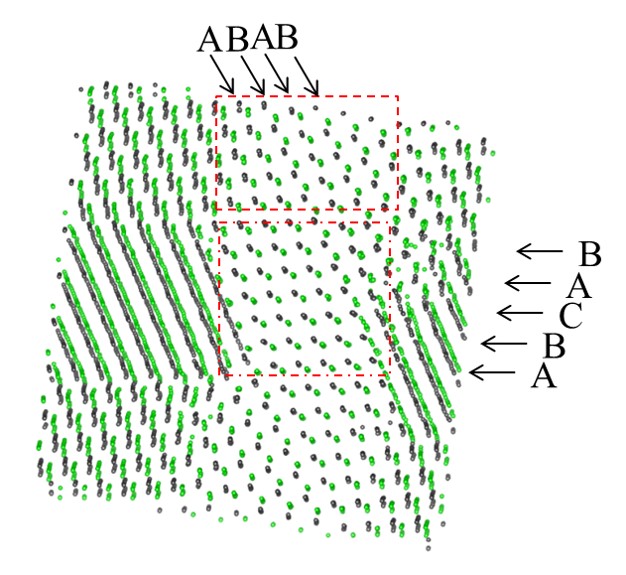}%
 \caption{\label{FIG.7.}The atomic configurations for the system at $n$=40 K and $T$=50 K for DL potential. ABAB$\cdots$ atomic arrangements are circled by dashed lines, and ABCABC$\cdots$ arrangements by dotted lines, where A, B, and C are the close packed atomic layers. S$_{1}$ and S$_{2}$ Si atoms are indicated by black and green dots, respectively.}
 \end{figure}
\section{Results and Discussions}
\subsection{The phase transition and identification of the crystal structure}
We firstly calculated the energies and volumes of the Si atom at different temperatures based on DL potential and other interatomic potentials proposed previously including Tersoff, MEAM, EDI, SW, and COMB potentials. There are also some modified versions for these potentials, for example, Tersoff-Mod (TM) potential \cite{Kumagai-1} and Tersoff-ModC (TMC) potential \cite{Pun-1} from Tersoff model, and MEAM-Spline (MEAMS) potential \cite{Lenosky-1} from MEAM. Though TMC potential is a newly developed one and has showed great improvement in comparison with other potentials, the simulated results are similar to those of TM potential and will not be presented. Figure 1 shows the dependences of energy per atom and volume per atom on the temperature. As indicated in Fig. 1(a), For the Tersoff potential case, the system is always in a simple crystalline state. The simulation fails because the volume of the system quickly increases to infinity once the temperature is above 1800 K (Fig. 1(a)). This means that the system with Tersoff potential will never exhibits a liquid state and always shows the diamond structure. In the case of TM potential, the liquid state can be obtained. The energy per atom deceases with the decrease of temperature until the temperature approaches about 1300 K, where there is an abrupt change in energy, indicating a phase transition. At the same time, there is a sudden increase in the volume, and this happens also for the simulations with EDI, MEAM, MEAMS, SW, and COMB potentials (Fig.1(b)). In addition, both Fig.1(a) and Fig.(b) show a clear thermal hysteresis, especially for the simulation by DL potential. Also for the DL potential case, both the energy and volume decrease when the system is cooling from high temperature, and there is a liquid-crystalline phase transition. The energy per atom in the case of DL potential is about -2.18 eV at 100 K, only about half of those from other potentials. However, the energy can still ensure a crystallization temperature comparable to that manifested by experiment (see Table II). The value for the volume is also in good agreement with experimental data (see Table II).

Figures 2 and 3 show the distribution functions of the distances between atoms $\rho(d)$  and the angles between the lines linking an atom with its nearest neighbors $\rho(\theta)$  simulated with different interatomic potentials in MD simulations at $T$=100 K and 4000 K, respectively. In Fig. 2, $\rho(d)$  and $\rho(\theta)$  for diamond structure are not zero for some certain $d$ and $\theta$. When the system is cooled down, $\rho(d)$  and $\rho(\theta)$  for Tersoff potential match quite well with those from perfect Si crystal, indicating that Tersoff potential is good for keeping the diamond structure stable. However, the system with TM potential shows a clear change. For example, $\rho(\theta)$  has a weak peak at 109.5$^{\circ}$  at $T$=100 K, but it is not crystalline from $\rho(d)$  as shown in Fig. 2(a). At $T$=4000 K, these peaks disappear, and all the systems are in a liquid state (see Fig. 3(a)). 

Figure 4 shows the atomic configurations with different interatomic potentials applied in MD simulations at $T$=100 K. In Fig. 4(h), S$_{1}$ and S$_{2}$ atoms are indicated by black and green dots, respectively. As shown in Figs. 4(e) and 4(f), the system with Tersoff potentials is still crystalline, and the one with TM potential is clearly disordered. The abnormal change in energy may mean a transition from one liquid phase to another non-crystalline phase, as also indicated in Fig. 1. The behaviors of the systems with EDI, MEAM, MEAMS, SW, and COMB potentials are also similar to that for TM potential (see Figs. 2, 3, and 4). Therefore, it can be concluded from Figs. 1, 2, 3, and 4 that with MEAM, MEAMS, SW, EDI, TM, and COMB potentials the diamond structure could not be reproduced in MD simulations and there are no lattice constants.

In addition, in Fig. 2, with DL potential, $\rho(d)$  and $\rho(\theta)$  are in agreement with the data from the reference diamond structure, even though the seventh peak of $\rho(d)$  shows a small deviation. Fig. 3 shows that $\rho(d)$  and $\rho(\theta)$  at $T$=3000 K are similar to those for MEAM, MEAMS, SW, EDI, TM, and COMB potentials. Fig. 4(h) indicates that every green atom has four black nearest atoms, and every black atom has also four green nearest atoms. However, the calculated results showed that the wurtzite structure has the same $\rho(d)$  and $\rho(\theta)$  as the diamond structure, so we could not tell the exact crystal structure from $\rho(d)$  and $\rho(\theta)$ . In order to identify the crystal structure, we need to check the crystal lattice of the subsystem. It is known that for the diamond structure, the subsystem shows the fcc lattice, and for the wurtzite structure, the subsystem shows the hcp lattice. 

\subsection{The relation between the diamond structure and fcc sublattice or hcp sublattice}
Previous investigation has shown that the ground state for LJ potential corresponds to the hcp lattice \cite{Zhang-1}. However, because the energy for fcc lattice is slightly higher than that for hcp lattice \cite{Kihara-1,Barron-1}, with LJ potential the system can also show the fcc lattice under some simulation conditions \cite{Zhang-1}. In this work, by changing the temperature step of the cooling for the system $n$, we can investigate the effect of $n$ on the crystal structure of the system. Figure 5 shows the dependences of energy per atom and volume per atom on the temperature with different temperature steps $n$ in MD simulations for DL potential. For $n$=5 K the system shows the lowest energy and volume, and the highest crystallization temperature of 1780 K. The volume increases slightly with increasing temperature step.

Figure 6 shows the distribution functions of $\rho(d)$  and $\rho(\theta)$  with different temperature steps $n$ applied in MD simulations at low temperatures for DL potential. In Figs. 6(c) and 6(e), for S$_{1}$ (or S$_{2}$) subsystem, the lattice constant of the crystal cell is $a_{f}$, which corresponds to the $d_{2}$ value at which the subsystem shows the second peak in $\rho(d)$ . The $d_{1}$ value is the shortest distance between atoms. From $\rho(d)$  for the diamond structure, the lattice constant for the diamond cell $a_{d}$ is equal to $a_{f}$. There are four distances in one diamond cell. $d_{2}$ and $d_{4}$ are accordingly $d_{1}$ and $d_{2}$ for subsystems, respectively, and $d_{1}$ and $d_{3}$ are the distances between S$_{1}$ and S$_{2}$ atoms, respectively. $d_{1}$ and $d_{2}$ have been defined in LJ potentials, but the distances between S$_{1}$ and S$_{2}$ atoms from the simulations are slightly different from those defined in LJ potentials. The angles are 60$^{\circ}$, 90$^{\circ}$, and 120$^{\circ}$  for fcc sublattice, and 60$^{\circ}$, 90$^{\circ}$, 109.5$^{\circ}$, and 120$^{\circ}$ for hcp sublattice. 

At $n$=5 K, both S$_{1}$ and S$_{2}$ subsystems show the hcp lattice for the third $d$ peaks and $\theta$=109.5$^{\circ}$ peaks coexist in comparison with other cooling rates (see Fig. 6).  However, there are also a number of atoms in S$_{1}$ and S$_{2}$ subsystems showing fcc lattice when the temperature step increases. For instance, at $n$=40 K, during the heating, there is a transition of fcc phase to hcp phase (Fig. 5(a)). Not all atoms in the simulation box are showing the hcp lattice, and instead a mixture of fcc and hcp lattices coexists. 

Figure 7 shows the atomic configurations for the system at $n$=40 K and $T$=50 K for DL potential. S$_{1}$ and S$_{2}$ Si atoms are indicated by black and green dots, respectively. The atomic arrangement for the fcc lattice is ABCABC$\cdots$, and the hcp lattice shows ABAB$\cdots$ arrangement, where A, B, and C are the close packed atomic layers, as indicated by dotted and dashed circles in Fig. 7. At $n$=100 K, the system is checked to mainly show the diamond structure with clear misalignment. As indicated in Figs. 5 and 6, the crystallization temperature (the temperature for the transition of liquid phase to crystalline phase) is 1602 K, and the lattice constant is 5.209 \AA, comparable with its experimental values of 1687 K and 5.430 \AA \cite{Kittel-1}.

Both the wurtzite structure and diamond structure have the same energy \cite{Pun-1}, and the ground state for LJ potential is the hcp lattice. Therefore, the ground state for DL potential is the wurtzite structure. In practice,  the diamond structure can be reproduced for higher cooling rates because the system needs a long simulation time to form the hcp lattice.  

To summarize, the system with DL potential exhibits the wurtzite structure for very slow cooling rates and the diamond structure for rapid cooling rates. If the interatomic potential with its ground state as the fcc lattice is used, the diamond structure will be formed and the investigation is still on the way.

\subsection{A comparison of DL model with experiment results}
\begin{table*}
\caption{The crystallization temperature $T_{c}$, lattice constant $a$, and elastic constants ($c_{11}$, $c_{12}$, and $c_{44}$) obtained with different interatomic potentials applied in MD simulations.}
\begin{tabular}{ p{20mm}p{20mm}p{20mm}p{20mm}p{20mm}p{20mm}p{20mm}}
\hline
\toprule
Potential & 	$T_{c}$(K)	& $a$(\AA) &  $c_{11}$(GPa) & $c_{12}$(GPa) & $c_{44}$(GPa)\\
\hline
\midrule
EDI	 & & &165.83 & 69.86 & 112.09\\
MEAM & & & 157.08 & 63.44 & 194.65\\
MEAMS & & & 113.24 & 62.45 & 86.72\\
SW & & & 149.37 & 75.91  & 108.36\\
Tersoff & & & 120.62  & 82.33  & 85.24\\
TM & & & 154.59 & 68.41 & 116.17\\
COMB & & & 138.45 & 73.17 & 113.78\\
DL($n$=100 K) & 1602 & 5.209 &  187.26 & 106.22 & 106.03\\
EXP\cite{Kittel-1} & 1687 & 5.430 &  165.78 & 63.97 & 79.62\\
\bottomrule
\hline
\end{tabular}
\end{table*}
Table II lists the crystallization temperature, lattice constant, and elastic constants obtained with different interatomic potentials applied in MD simulations. It is shown that only the simulation based on DL potential gives the diamond structure. For other potentials, the simulations have revealed that the systems could not form the diamond structure. In addition, for DL potential, the elastic constant $c_{11}$  is about 13$\%$ larger than the experimental data, and $c_{12}$ =$c_{44}$. Despite this, in contrast with the results from other potentials, the agreement between the simulated results and the experimental data proves the validity of DL potential. The present result indicates that the crystal structure should be the first criterion for the validity test for the interatomic potentials.

LJ potential has been employed in MD simulations since several decades before, and is thought to be only suitable for rare gas or close packed atom systems by most researchers \cite{Kihara-1,Barron-1,Kittel-1}. With the crystal structure as the first criterion for the validity test, our previous results have demonstrated that with one single LJ potential one can reproduce only the hcp or fcc lattice, and never reproduce other Bravais lattices such as body centered cubic (bcc) lattice \cite{Zhang-1,Zhang-4}. With the combination of several LJ potentials and without setting any initial Bravais lattices, one can obtain the diamond structure and graphite structure \cite{Zhang-2}, CsCl and NaCl structures \cite{Zhang-4}, perovskite (ABO$_{3}$) structure \cite{Zhang-3}. This demonstrates that LJ potential is of significant importance in the construction of interatomic potentials.

For Tersoff and EDI potentials, except the pair term, three-body term has been introduced to define the bond angle and actually shows its power in the formation of the bond angle of 109.5$^{\circ}$ in MD simulations, as shown in Fig. 2(b). Unfortunately, the introduction of a three-body term does not lead to the diamond structure. Similarly, the COMB potential cannot give the diamond structure even when the charge is taken into account.

In order to reproduce the diamond structure in MD simulations with DL potential, three requirements must be satisfied. First, two types of atoms must be created even though they are physically identical to each other. Second, these two types of atoms must show their own sublattices. Here, S$_{1}$ and S$_{2}$ atoms can show the hcp lattice, or the fcc lattice. Third, the distance between S$_{1}$ and S$_{2}$ atoms determines the resulting crystal structure. If the shortest distance for S$_{1}$ (S$_{2}$) atoms is $d_{A}$, and the shortest distance between S$_{1}$ and S$_{2}$ atoms is $d_{AB}$. When the ratio of $d_{A}/d_{AB}$ is in the range of 1.2-1.3, one can obtain CsCl structure. $d_{A}/d_{AB}$ values lie in the range of 1.3-1.6 for NaCl structure, 1.7-1.9 for the diamond structure, and 1.9-2.1 for graphite structure. For graphite structure, S$_{1}$ and S$_{2}$ atoms may show the hcp or fcc lattices, leading to $\alpha$ graphite or $\beta$ graphite, respectively. In the graphite structure, every single layer is a graphene.

For LJ potential, it is very interesting that, even though the ground state corresponds to the hcp lattice, the lattice for the subsystem can be self-adaptive to form an ordered structure when the distance between S$_{1}$ and S$_{2}$ atoms changes. In the case of CsCl structure, the sublattice is the simple cubic (sc) lattice with $d_{A}/d_{AB}$ =1.2, and for NaCl structure, sublattice is only the fcc lattice with $d_{A}/d_{AB}$ =1.4. This means that in MD simulations with LJ potentials, the systems can self-adapt its sublattice to form an energetically favorable ordered structure. However, because LJ potential shows a spherical symmetry, the resulting self-adaptive sublattice also shows the spherical symmetry.

\section{Conclusions}
Here we have proposed the DL potential for MD simulation of silicon and the validity of DL potential and other potentials has been tested. The results indicate that with TM, SW, EDI, COMB, MEAM, and MEAMS potentials the diamond structure cannot be reproduced in MD simulations. Only the system with DL potential can show the diamond structure. The crystallization temperature, lattice constant, and elastic constants have been obtained from the simulations with DL potential, and the results are in agreement with experimental data.

%
%
%

\begin{acknowledgments}
Hui Zhang thanks M. P. Allen, G. P. Pun, and G. J. Ackland for useful discussions.This work is supported by the National Natural Science Foundation of China (Grant No. 11204087).
\end{acknowledgments}

\end{document}